# A Weighted U Statistic for Genetic Association Analyses of Sequencing Data

Changshuai Wei,[1,2] Ming Li,[3] Zihuai He,[4] Olga Vsevolozhskaya,[1] Daniel J. Schaid,[5] and Qing Lu[1] *

[1]*Department of Epidemiology and Biostatistics, Michigan State University, East Lansing, Michigan, United States of America;*

[2]*Department of Biostatistics and Epidemiology, University of North Texas Health Science Center, Fort Worth, Texas, United States of America;*

[3]*Division of Biostatistics, Department of Pediatrics, University of Arkansas for Medical Sciences, Little Rock, Arkansas, United States of America;*

[4]*Department of Biostatistics, University of Michigan, Ann Arbor, Michigan, United States of America;*

[5]*Division of Biomedical Statistics and Informatics, Mayo Clinic, Rochester, Minnesota, United States of America*

Running title: A Weighted U for Sequencing Data Analyses

*Correspondence to:
Qing Lu, Ph.D
Department of Epidemiology and Biostatistics, Michigan State University;
B601 West Fee Hall, East Lansing, MI 48824
Phone: 517.353.8623 x137
Fax: 517.432.1130
E-mail: qlu@epi.msu.edu

# Abstract

With advancements in next generation sequencing technology, a massive amount of sequencing data are generated, offering a great opportunity to comprehensively investigate the role of rare variants in the genetic etiology of complex diseases. Nevertheless, this poses a great challenge for the statistical analysis of high-dimensional sequencing data. The association analyses based on traditional statistical methods suffer substantial power loss because of the low frequency of genetic variants and the extremely high dimensionality of the data. We developed a weighted U statistic, referred to as WU-SEQ, for the high-dimensional association analysis of sequencing data. Based on a non-parametric U statistic, WU-SEQ makes no assumption of the underlying disease model and phenotype distribution, and can be applied to a variety of phenotypes. Through simulation studies and an empirical study, we showed that WU-SEQ outperformed a commonly used SKAT method when the underlying assumptions were violated (e.g., the phenotype followed a heavy-tailed distribution). Even when the assumptions were satisfied, WU-SEQ still attained comparable performance to SKAT. Finally, we applied WU-SEQ to sequencing data from the Dallas Heart Study (DHS), and detected an association between *ANGPTL 4* and very low density lipoprotein cholesterol.

# Introduction

Genome-wide association studies (GWAS) have been used to uncover common genetic variants predisposing to common complex diseases. During the past decade, thousands of new genetic variants have been identified from GWAS, some with compelling biological plausibility for a role in disease etiology [Hindorff, et al. 2009]. Despite such successes, the genetic variants identified for most complex diseases thus far can only explain a small fraction of the total heritability. Evolutionary theory suggests that rare variants are likely to be recent mutations, and are also likely to be more deleterious because they are under less negative selective pressure [Boyko, et al. 2008; Fay, et al. 2001; Kryukov, et al. 2007; Pritchard 2001; Raychaudhuri 2011]. Converging evidence from genetic studies of complex diseases also suggests that complex diseases are highly heterogeneous, and that familial subtypes of complex diseases are frequently seen to be caused by rare variants with large effects [Ahituv, et al. 2007; Cohen, et al. 2004; Easton, et al. 2007; Ji, et al. 2008; Romeo, et al. 2009]. Nevertheless, current findings of rare variants predisposing to complex diseases are still limited. With advancements in high-throughput sequencing technology, researchers are now able to comprehensively study the role of a massive amount of sequencing variations, including both common and rare variants. While the ongoing exome and whole-genome sequencing studies hold great promise for identifying new genetic variants, including rare variants, they also pose great challenges for the statistical analysis of massive amounts of sequencing data.

Although single-locus analysis has been widely used in GWAS, such analysis is underpowered for detecting rare variants. Rare variants are anticipated to have larger effects than common variants, but their low frequencies in the study population make them hard to detect. Moreover, the number of single-locus tests required for sequencing studies is significantly larger than those required for GWAS studies, which increases the multiple-testing burden. Therefore, as an essential alternative, a joint association analysis of genetic variants, became popular in the association analysis of sequencing data [Barnett, et al. 2013; Chen, et al. 2012; Ladouceur, et al. 2012; Lee, et al. 2012; Li and Leal 2008; Lin and Tang 2011; Madsen and Browning 2009; Morgenthaler and Thilly 2007; Neale, et al. 2011; Wei and Lu 2011; Wu, et al. 2011; Zhu, et al. 2010]. A common strategy used in such analysis is to first aggregate the effect of rare variants in a region (e.g., a gene) via collapsing or weighting, and then to assess the joint association of the variants with the phenotype of interest. By aggregating information from multiple variants, the association signal is enhanced and becomes more detectable. Moreover, this greatly reduces the number of tests, which alleviates the multiple-testing issue.

The commonly used methods for joint association analysis include the cohort allelic sum test (CAST), the combined multivariate and collapsing (CMC) method [Li and Leal 2008] and the weighted sum test (WST) [Madsen and Browning 2009]. These methods, however, implicitly assume the effects of different rare variants are in the same direction or magnitude. To address this limitation, the C-alpha test [Neale, et al. 2011] was proposed to consider the direction and magnitude of effects, whereby the expected variance and observed variance were compared in a case-control setting. Another widely used semi-parametric method for sequencing data analyses is the sequence kernel association test (SKAT) [Wu, et al. 2011], which is also robust for the direction and magnitude of genetic effects. Moreover, built based on the kernel machine regression, SKAT can adjust for covariates, and is applicable to both binary and Gaussian phenotypes.

Most existing methods for sequencing data analyses are parametric-based or semi-parametric based, which rely on certain assumptions. However, in practice, these assumptions may not be satisfied (e.g., the phenotype may not follow a normal distribution in a sequencing study based on an extreme phenotype design). When the assumptions are violated, the existing methods will likely have either decreased power or inflated type I error. Non-parametric methods, such as U-statistic-based methods, have shown their robustness against underlying phenotypic distributions and/or underlying modes of inheritance [Li 2012; Li, et al. 2011; Schaid, et al. 2005; Wei, et al. 2012; Wei, et al. 2013; Wei, et al. 2008]. Various U-statistic-based methods have been proposed to identify common variants associated with binary or quantitative phenotypes. Most of these methods use U statistic to obtain multiple group-wise scores, and then form test statistics to compare scores among different groups. For case-control data analyses, U statistic is used to summarize the genetic information, and then to compare scores between cases and controls [Schaid, et al. 2005]. For quantitative data analyses, U statistic first summarizes the phenotypic information, and then compares scores among different genotype groups or multi-locus genotype groups [Wei, et al. 2008]. Because of different ways of constructing U-statistics for binary and quantitative phenotypes, it is challenging to provide a unified method under the traditional U-statistic framework [Li 2012].

The weighted U statistic is a more general form of the U statistic. It was developed in the 1980s-1990s [Gregory 1977; Serfling 1981; Shieh 1997] and has rarely been used in genetic data analyses. By using a weighted U statistic, genetic information and phenotypic information can be summarized separately into the weight function and U kernel, thereby avoid the issue of group score comparison. Based on the weighted U statistic, we develop a unified method, referred to as WU-SEQ, for sequencing data analyses of various types of phenotypes (e.g., binary, ordinal, and continuous). Moreover, we use a projection method in WU-SEQ for covariates adjustment, and derive the asymptotic distribution of the test statistic for an efficient assessment of the significance of the association. The performance of WU-SEQ was evaluated though extensive simulation studies, and compared with a commonly used method, SKAT [Wu, et al. 2011]. Finally, we illustrated the proposed method by applying WU-SEQ to sequencing data from the Dallas Heart Study (DHS).

## Methods

### A weighted U for association analyses of sequencing data

Assume a population-based sequencing study with *N* unrelated subjects and *P* single nucleotide variants (SNV) located in a gene or a genetic region. Let $y_i$ and $G_i = (g_{i1}, g_{i2}, \cdots, g_{iP})$ denote, respectively, the phenotype and the genotypes of *P* variants of an individual *i* ($1 \leq i \leq n$), where $g_{ip}$ ($1 \leq p \leq P$) is coded as 0, 1, or 2. We use $s_{i,i'}$ and $w_{i,i'}$ to denote the phenotypic similarity and genetic similarity between individuals *i* and *i'*, respectively. The phenotypic similarity can be measured by any 2 degree kernel function, $s_{i,i'} = h(y_i, y_{i'})$, which satisfies the finite second moment condition ($E_F(h^2(Y_1, Y_2)) < \infty$). While various kernel functions can be used to measure the phenotypic similarities, in this paper, we use the cross product kernel, $s_{i,i'} = q_i q_{i'}$, where $q_i$ is the normal quantile of the rank of $y_i$, defined as $q_i = \Phi^{-1}((rank(y_i) - 0.5)/n)$, and $\Phi^{-1}(\bullet)$ is the inverse cumulative distribution function for standard normal distribution. In the presence of ties, we assign them an averaged rank. For example, if there are $n_0$ numbers of controls (i.e., $y_i = 0$) and $n_1$ numbers of cases (i.e., $y_i = 1$), an average rank is assigned to the group with the same phenotype value (e.g., $rank(y_i \in \{y_i, y_i = 0\}) = (n_0 + 1)/2$ is assigned to the control group). Other than the quantile-transformed cross product kernel, distance-transformed phenotypic similarities, such as $s_{i,i'} = \exp(-|y_i - y_{i'}|)$ or $s_{i,i'} = \exp(-(y_i - y_{i'})^2)$ can be centralized and be used to measure the phenotypic similarity. Nevertheless, as we demonstrate below, the use of the quantile-transformed cross product kernel leads to nice asymptotic properties. Genetic similarity, $w_{i,i'}$, can be calculated using a variety of similarity functions. One of the commonly used similarity functions for sequencing data is the weighted IBS, which gives more weight to rare variants,

$$w_{i,i'} = \sum_{p=1}^{P} \frac{2-|G_{i,p} - G_{i',p}|}{\Upsilon \sqrt{\gamma_p (1-\gamma_p)}},$$

where $\gamma_p$ is the minor allele frequency for the $p$-th rare variant, and $\Upsilon = \sum_{p=1}^{P} 2/\sqrt{\gamma_p (1-\gamma_p)}$ is used to standardize the weight function so that $w_{i,i'} \in [0,1]$. In addition to weighted IBS, distance-transformed similarity functions can also be used. For example, we could use $w_{i,i'} = \exp(-D_{i,i'})$, where $D_{i,i'}$ is the distance function (e.g., Euclidian distance).

Given $s_{i,i'}$ and $w_{i,i'}$, the weighted U is formed to measure the association of $P$ genetic variants with the disease phenotype,

$$U_w = \frac{1}{n(n-1)} \sum_{i \neq i'} w_{i,i'} s_{i,i'}, \tag{1}$$

where $s_{i,i'}$ is the 2 degree U kernel and $w_{i,i'}$ is the weight function for the weighted U. When $w_{i,i'} \equiv 1$, we can construct an un-weighted U by using only the phenotype similarity,

$$U_{uw} = \frac{1}{n(n-1)} \sum_{i \neq i'} s_{i,i'}. \tag{2}$$

In the weighted U, the summation is over all phenotypic similarities weighted by the genotypic similarity, whereas only the phenotypic similarity is used for the un-weighted U. Note that the U kernel, $s_{i,i'}$, could be a positive/negative value with a mean of 0. When there is no association (i.e., under the null), both the un-weighted U and the weighted U have an expectation of 0. When the genetic variants are associated with the phenotype, we would expect genetic similarities to be concordant with phenotypic similarities, where larger weights are given to larger values of phenotypic similarity. Therefore, the weighted U is expected to have a positive value, while the un-weighted U remains have an expectation of 0 (i.e., the weighted U is expected to be greater than the un-weighted U). Based on this concept, we could build an association test by comparing the weighted U with the un-weighted U. The two U statistics, however, are based on weights of different scales (i.e., $w_{i,i'}$ vs. constant 1), therefore, a constant $c$ is introduced to balance the two weight functions. The test statistic is then defined as,

$$WU_{seq} = U_w - cU_{uw}, \qquad (3)$$

where the scaling constant $c$ can be obtained by minimizing the L2 norm distance between the two weight metrics, i.e., $c = \arg\min_{c>0}\{\sum_{i \neq i'}(w_{i,i'} - c)^2\}$. Alternatively, we could choose other types of $c$. For instance, we could obtain $c$ by minimizing the L1 norm distance between the two weight metrics, i.e., $c = \arg\min_{c>0}\{\sum_{i \neq i'}|w_{i,i'} - c|\}$.

**Asymptotic distribution of the test statistic**

When the weighted U is significantly larger than the rescaled un-weighted U, we reject the null hypothesis and conclude there is an association between $P$ genetic variants and the phenotype. The p-value can be obtained by comparing the observed test statistic, $WU_{seq}^{obs}$, with the null distribution, i.e., $\Pr(WU_{seq} > WU_{seq}^{obs})$. For a small sample size, a permutation test can be used for the calculation of the p-value. However, for a large sample size and high-dimensional data, a permutation test could be computationally intense. Therefore, we derive the asymptotic distribution of $WU_{seq}$ to efficiently assess the significance level of the association.

We first rewrite the test statistic, $WU_{seq}$, of equation (3) as a weighted summation of $s_{i,i'}$,

$$WU_{seq} = \frac{1}{n(n-1)}\sum_{i \neq i'} k_{i,i'} s_{i,i'}, \qquad (4)$$

where we define a new weight $k_{i,i'} = w_{i,i'} - c$. Denote $Q = (q_1, q_2, ...q_N)^T$ and $K = \{k_{i,i'}\}_{n \times n}$. If the phenotypic similarity is measured by the cross product kernel, $s_{i,i'} = q_i q_{i'}$, $WU_{seq}$ is simplified to a quadratic form $n(n-1)WU_{seq} = \sum_{i \neq i'} q_i k_{i,i'} q_{i'} = Q^T K Q$, with all the diagonal elements of $K$ equal to 0 ($k_{i,i} = 0$). In such a case, it has a close connection with the variance component score test in the linear mixed model, except that $WU_{seq}$ does not use information from the diagonal terms ($k_{i,i} = 0$), and does not assume a Gaussian distribution of the phenotype.

The limiting distribution of U depends on $\zeta_1 = Var(E(h(Y_1, Y_2)|Y_2))$ [Serfling 1981]. If $\zeta_1 > 0$, the U statistic is a non-degenerated U and asymptotically follows a normal distribution. If $\zeta_1 = 0$, the U statistic is a degenerated U and can be approximated by a mixture chi-squared distribution. By the definition of $WU_{seq}$, we have $E(q_1 q_2 | q_2) = q_2 E(q_1) = 0$, and therefore $\zeta_1 = 0$. Because $\zeta_1 = 0$, $WU_{seq}$ is a degenerated weighted U statistic. Its limiting distribution can be approximated as a linear combination of chi-squared random variables,

$$nWU_{seq} \sim \sum_{m=1}^{\infty} \alpha_m \sum_{l=1}^{n} \tilde{\lambda}_l (\chi^2_{1,ml} - 1), \quad (5)$$

where $\chi^2_{1,ml}$ are iid chi-squared random variables with 1 degree of freedom. $\tilde{\lambda}_l$ and $\alpha_m$ are generated from the eigen-decomposition of the weight function $k_{i,i'}$ and the kernel function $s_{i,i'}$ [Serfling 1981; Shieh, et al. 1994; Wet and Venter 1973]. $\tilde{\lambda}_l$ ($l=1,\ldots,n$) are obtained from the eigen-values $\lambda_l$ of matrix $K = \{k_{i,i'}\}_{n \times n}$, with $\tilde{\lambda}_l = \lambda_l / (n-1)$. $\{\alpha_m\}$ are the eigen-values of a general kernel function $h(\bullet, \bullet)$, obtained from the following decomposition,

$$h(q_1, q_2) = \sum_{m=1}^{\infty} \alpha_m \varphi_m(q_1) \varphi_m(q_2),$$

where $\{\varphi_m(\bullet)\}$ are the ortho-normal eigenfunctions corresponding to $\alpha_m$. For the cross product kernel, we can show that $h(q_1, q_2) = q_1 q_2 = 1 \varphi_1(q_1) \varphi_1(q_2) + \sum_{m=2}^{\infty} 0 \varphi_m(q_1) \varphi_m(q_2)$, where $\varphi_1(q) = q$ (Appendix S1). Thus, $\alpha_m = 1_{\{m=1\}}$, where $1$ is an indicator function, and $\sum_{m=1}^{\infty} \alpha_m \sum_{l=1}^{n} \tilde{\lambda}_l = \sum_{l=1}^{n} \tilde{\lambda}_l = trace(K)/(n-1) = 0$. The limiting distribution of $WU_{seq}$ can be simplified to $nWU_{seq} \xrightarrow{D} \chi^2_{\Sigma,\infty}$, where $\chi^2_{\Sigma,\infty} = \sum_{l=1}^{\infty} \tilde{\lambda}_l \chi^2_{1,l}$ is a mixture chi-squared distribution with mean 0 and finite variance (Appendix A). Given the asymptotical distribution of $WU_{seq}$, the p-value can then be calculated using the Davis method [Davies 1980].

**Adjusting covariates with confounding effects**

Assume we have $J$ covariates, $X_i = (1, x_{i1}, \cdots, x_{iJ})$, $i = 1, 2, ..., N$. To adjust for the potential confounding effects, we will first fit the transformed value of the phenotype, $Q = (q_1, q_2, ... q_N)^T$, with covariates $X = (X_1, X_2, ..., X_N)^T$, and then use the residuals for the association test. Based on this idea, we project $Q$ onto the space spanned by $X$, and obtain the prediction $\hat{Q}$, where $\hat{Q} = X(X^T X)^{-1} X^T Q$. Denoting $H = I - X(X^T X)^{-1} X^T$ and $\hat{\sigma}^2 = (Q - \hat{Q})^T (Q - \hat{Q}) / (N - J - 1)$, we can obtain the residuals, $Q^e = HQ / \hat{\sigma}$. It's easily to show that $Q^e \perp X$, therefore, a new response vector is attained that is perpendicular to the space spanned by the covariates. The test statistic $WU_{seq}$ can be reconstructed as

$$WU_{seq} = \frac{1}{n(n-1)} \sum_{i \neq i'} k_{i,i'} q_i^e q_{i'}^e.$$

Using the same argument as above, $WU_{seq}$ with covariates adjustment can also be approximated by a linear combination of chi-squared random variables,

$$nWU_{seq} \sim \sum_{l=1}^{n} \tilde{\lambda}_l^e \chi_{1,l}^2.$$

$\tilde{\lambda}_l^e = \lambda_l^e / (n-1)$ and $\{\lambda_l^e\}$ are the eigen-values of matrix $K^e$, where $K^e = HKH$. If the model is correctly specified and the covariates (e.g., principle components) can capture the confounding effects, the residual confounding is negligible. However, in practical, we should consider potential issues, such as nonlinearity of covariates and high correlations between genetic variables and covariates, before implementing the covariate adjusting approach. Directly using the approach without considering these issues may lead to inflated type I error or power loss. For instance, principle components from genetic data can be used to adjust for confounding effects due to population stratification or population admixture. Nevertheless, if the genetic variants and/or the number of principle components are insufficient, residual confounding could lead to inflated type I error[Price, et al. 2006].

## Results

## Simulation

We conducted simulation studies to evaluate the performance of WU-SEQ, comparing it with one of the most commonly used methods, SKAT [Wu, et al. 2011]. For all of the simulations, we used genetic data from the 1000 genome project [Abecasis, et al. 2010] to mimic the real sequencing data structure (e.g., LD pattern and allele frequency). Specifically, we used a 1Mb region from the genome (Chromosome 17: 7344328-8344327), and randomly chose a 30kb segment from that 1Mb region for each simulation replicate (if not specified otherwise). The average number of SNVs in the sampled 30kb segments is 194. The minor allele frequency (MAF) of the SNV in the genome region ranged from $4.50 \times 10^{-4}$ to $4.99 \times 10^{-1}$, with a distribution highly skewed to rare variants (34.8% of the variants with MAF<0.001, 69.1% of the variants with MAF<0.01 and 80% of the variants with MAF<0.03). Similar to the SKAT simulation studies [Wu, et al. 2011], we selected a portion of the genetic variants with MAF<0.03 as functional SNVs. A number of individuals, ranging from 50 to 500, were randomly chosen as the study samples from all of the available individuals in the 1000 genome project.

For each disease model, we simulated 1000 data replicates and obtained the type 1 error and power for both WU-SEQ and SKAT. For SKAT, we used the link function according to the distributions of phenotype (i.e., the logit link for a binary phenotype and the identity link for a continuous phenotype) [Wu, et al. 2011]; for WU-SEQ, we used the L2 norm to choose the constant $c$ and did not specify an assumption on the phenotype distribution. To be consistent, both methods used a weighted IBS to summarize genetic information. The type 1 error and power were obtained by calculating the percentage of p-values smaller than 0.05 from 1000 data replicates.

*Simulation I*

We first simulated a series of disease models, without considering covariates, and investigated the influence of the direction and magnitude of effects on both methods. Each data replicate was comprised of 500 subjects. We considered 4 types of distributions for the phenotypes: binary, Gaussian, Student's t with 2 degrees of freedom, and Cauchy. Binary and Gaussian phenotypes are typically observed in population-based studies. The Cauchy-distributed and t-distributed phenotypes represent continuous phenotypes with more extreme values (i.e., heavy tailed). We used the logistic model to simulate the binary phenotype,

$$\text{logit}(P(y_i = 1)) = \mu + G_i \beta,$$

where $G_i$ and $y_i$ were the genotype and phenotype of the $i$-th individual, respectively. $\beta$ was a vector of regression parameters, measuring the effects of the genetic variants. For each simulation replicate, we sampled an effect vector from a multivariate normal distribution, $MVN(\mu_\beta \vec{1}, \sigma_\beta^2 I)$, where $\vec{1}$ was the vector of 1 and $I$ was the identity matrix. For Gaussian phenotypes, we simulated the model as,

$$y_i = \mu + G_i\beta + \varepsilon_i,$$

where $\varepsilon_i \sim N(0, \sigma^2)$. For the t-distributed phenotype, we simulated the model as:

$$y_i = \mu + G_i\beta + \varepsilon_i, \varepsilon_i \sim t_{df=2}.$$

For the Cauchy type of phenotype, we used

$$y_i \sim cauchy(a_i, b).$$

$a_i$ and $b$ were the location parameter and the scale parameter of the Cauchy distribution, respectively, where $a_i = \mu + G_i\beta$ and $b$ was a fixed value. For all four types of phenotypes, we considered different directions of genetic effects. For the first scenario, we assumed $\mu_\beta = 0$, whereby half of the functional SNVs were deleterious and half of the functional SNVs were protective. For the second scenario, we assumed $\mu_\beta > 0$, whereby the majority of the functional SNVs were deleterious. For each scenario, we varied the percentage of functional SNVs from 5% to 50%. The details of the simulation setting are provided in Table S1.

We summarized the results in Table 1. From Table 1, we can see that WU-SEQ had a well-controlled type 1 error rate under various phenotype distributions. In contrast, SKAT had an inflated type 1 error rate when the underlying distribution was the Cauchy distribution (0.194) or t distribution (0.110). Similar to SKAT, WU-SEQ allows for different directions of genetic effects (i.e., both deleterious and protective effects). For both scenarios (i.e., $\mu_\beta = 0$ and $\mu_\beta > 0$), WU-SEQ obtained a power that was comparative to or a slightly higher power than SKAT. As the percentage of the functional SNVs increased, both WU-SEQ and SKAT gained improved power for binary and Gaussian phenotypes. For the Cauchy phenotype and t-distributed phenotype, however, the power of WU-SEQ was significantly higher than that of SKAT. With the Cauchy phenotype, for example, SKAT's power ranged from 0.167 to 0.192 for $\mu_\beta = 0$ and ranged from 0.175 to 0.199 for $\mu_\beta > 0$, while the power of WU-SEQ increased from 0.116 to 0.635 for $\mu_\beta = 0$ and increased from 0.095 to 0.818 for $\mu_\beta > 0$.

*Simulation II*

In simulation II, we investigated the influence of sample size on the performance of WU-SEQ and SKAT. We used the same models as in simulation I to simulate binary, Gaussian, Student's t and Cauchy phenotypes. For simplicity, we assumed $\mu_\beta = 0$ and a fixed $\sigma_\beta^2$, and varied the sample size from 50 to 500 (Table S2). We assumed 0% of the genetic variants were functional for the null model, and assumed 50% of the genetic variants were functional for the disease models.

The results were summarized in Table 2. Type 1 error rates of WU-SEQ were well controlled at the 0.05 level for different sample sizes (i.e., 50, 100, 200 and 500) and various phenotype distributions (i.e., binary, Gaussian, Student's t and Cauchy), while the type 1 error of SKAT was inflated for the Cauchy phenotype (0.118~0.194) and the t-distributed phenotype (0.110~0.131). We also observed that both WU-SEQ and SKAT had conservative type I error rates when the study sample sizes were small. For instance, the type I error rates of SKAT and WU-SEQ were 0.001 and 0.005, respectively, when the sample size was 50 and the phenotype was binary. The power of WU-SEQ increased as the sample size increased for all four types of phenotypes. The power of SKAT remained almost the same for the Cauchy phenotype (0.126~0.177) when the sample size increased. For both the binary and Gaussian phenotypes, the power of WU-SEQ was comparable or slightly higher than that of SKAT, while the power of WU-SEQ was significantly higher than that of SKAT for the t-distributed and Cauchy-distributed phenotypes.

Additional simulations were also conducted to evaluate the performance of WU-SEQ in the high-dimensional data setting. Instead of using a 30kb segment, we randomly chose a 500kb segment for each simulation replicate. The average number of SNV for each subject was increased from 194 to 1873. We fixed the sample sizes as 100 and simulated the phenotype by using the settings presented in Table S2. The result showed that WU-SEQ had well controlled type 1 error and good power performance when the number of SNVs was larger than the sample size (Table 3).

*Simulation III*

In genetic association analyses, we often need to adjust important covariates (e.g., gender) for potential confounding effects. Therefore, we conducted another set of simulations to investigate the performance of WU-SEQ with the consideration of covariates adjustment. In this simulation, we simulated two covariates, $x_{1,i} \sim Bernoulli(0.3)$ and $x_{2,i} \sim N(0,1)$ for each subject. The binary phenotype was simulated by using

$$\text{logit}(P(y_i = 1)) = \mu + X_i\alpha + G_i\beta,$$

where $X_i = (x_{1,i}, x_{2,i})$ and $\alpha = (\alpha_1, \alpha_2)^T$ were the covariates and the effects of the covariates, respectively. Similarly, we used the linear model for the Gaussian phenotype,

$$y_i = \mu + X_i\alpha + G_i\beta + \varepsilon_i, \varepsilon_i \sim N(0, \sigma^2),$$

and the following model for the t-distributed phenotype,

$$y_i = \mu + X_i\alpha + G_i\beta + \varepsilon_i, \varepsilon_i \sim t_{df=2}.$$

The Cauchy phenotype was simulated by using,

$$y_i \sim cauchy(a_i, b), \text{ where } a_i = \mu + X_i\alpha + G_i\beta.$$

We varied the sample size from 50 to 500, and fixed the percentage of functional rare variants at 50% for the disease models (Table S3).

The results were summarized in Table 4. We found that the projection method worked very well for WU-SEQ in terms of covariate adjustment, and the type I error was well-controlled under different phenotype distributions. SKAT, however, had a well-controlled type 1 error only when the underlying assumptions were satisfied and the link function was correctly specified. Similar to simulations I and II, we observed comparable power between WU-SEQ and SKAT for both the binary and Gaussian phenotypes, and WU-SEQ had significantly higher power than SKAT for the Cauchy and t-distributed phenotypes. We also found the type 1 error rates were less conservative for studies with the binary phenotype with covariate adjustment. This could be due to the fact that the increased levels of residuals after covariate adjustment lead to a better approximation of normal distribution.

In addition to the above simulations, we conducted simulations to investigate the performance of WU-SEQ under different choices of $c$ (Table 5) and different choices of U kernel (Table 6). For different $c$ based on L1 norm and L2 norm, WU-SEQ had well-controlled type I error. However, WU-SEQ based on L2 norm had slightly higher power than that based on L1 norm (Table 5). We also compared the U kernel with the qunatile transformation and that without the quantile transformation (by using ranks of phenotypes). Both approaches could control the type I error. Nevertheless, using the quantile transformation let to higher power for normal and t distributed phenotypes (Table 6). In simulation studies, we used 1000 simulated data replicates to access the type I errors of WU-SEQ at 0.05 level. We also conducted additional simulation studies by using 100000 simulated data replicates to access the type I error of WU-SEQ at $5 \times 10^{-4}$ level. The results (Table S4) showed that the type I error at $5 \times 10^{-4}$ level is well controlled around $5 \times 10^{-4}$. Moreover, the type I errors at 0.05 level were much closer to 0.05 by using 100000 simulated data replicates.

**Application to the sequencing data from the Dallas Heart Study**

To further evaluate the performance of WU-SEQ and SKAT, we applied both methods to the sequencing data from the Dallas Heart Study (DHS) [Romeo, et al. 2009]. The DHS sequencing data is comprised of 4 genes, *ANGPTL3*, *ANGPTL4*, *ANGPTL5* and *ANGPTL6*. We were interested in evaluating the association of these genes with body mass index (BMI), cholesterol, and very low density lipoprotein cholesterol (VLDL). Following the previous study [Wu, et al. 2011], we considered age, gender, and race as covariates in the model. Prior to the association analysis, we re-assessed the quality of the genotype data. As a part of the quality assessment, we eliminated the SNVs and subjects that had a high missing rate. After the quality control, 230 rare variants (54 SNVs, 63 SNVs, 61 SNVs, and 52 SNVs are from *ANGPTL3*, *ANGPTL4*, *ANGPTL5* and *ANGPTL6*, respectively) and 2598 subjects remained for the analysis. In the analysis, random imputation based on allele frequency was used to impute missing genotypes.

The distribution of the SNVs in the DHS was heavily skewed to the rare variants (Figure S2), wherein 93.5%, 87.4%, and 70% of all SNVs had an MAF of less than 3%, 1% and 0.1%, respectively. Similar to previous studies [Wu, et al. 2011], we selected SNVs with an MAF<3% for the analysis in order to detect associations due to rare variants. The distributions of MAF for the 4 genes were given in Figure 1. From Figure 1, we can see that the distributions of MAF were highly skewed to the left, with the majority of the SNVs having an MAF<0.1%. The distributions of the three phenotypes, BMI, cholesterol and VLDL, can be viewed in Figure 2. Among the three phenotypes, the distribution of VLDL was heavily skewed to the left, which is unlikely to follow the Gaussian distribution. We applied both WU-SEQ and SKAT to the association analyses of 4 genes, with the consideration of three covariates: gender, race and age (Table 7). To be consistent, we used weighted IBS for both WU-SEQ and SKAT. WU-SEQ detected a strong association of *ANGPTL4* (p-value=0.007) with VLDL, while SKAT found only a marginal association (p-value=0.105). To further explore this finding, we combined all 4 genes into a gene set, and then tested its association with VLDL (Table 7). The association remained significant (p-value=0.032) using WU-SEQ, but not SKAT (p-value=0.316). None of the 4 genes was found to be associated with BMI using either WU-SEQ or SKAT. A marginal association was detected between *ANGPTL6* and cholesterol, with a p-value of 0.059 from WU-SEQ and a p-value of 0.025 from SKAT. To evaluate the performance of both methods for the binary phenotype, we dichotomized the data into a highest quartile (coded as 1), and a lowest quartile (coded as 0) for each phenotype (Table S5). Overall, WU-SEQ attained similar results to SKAT. For example, both methods found no association of all 4 genes with BMI, a marginal association of all 4 genes with cholesterol, and a significant association of all 4 genes with VLDL.

## Discussion

Targeted, exome and whole-genome sequencing studies are now underway for the discovery of new genetic variants, particularly rare variants, associated with complex diseases. With the emerging of a large amount of high-dimensional sequencing data, great challenges have been posed to statistical analyses of sequencing data. Conventional single-locus analyses have been shown to have low power for analyzing sequencing data, not only because of the low frequencies of rare variants, but also because of the use of a more stringent significance threshold. Joint association analyses of multiple genetic markers, as demonstrated in several studies, can greatly reduce the dimensionality and obtain powerful performance for sequencing data [Li and Leal 2008; Madsen and Browning 2009; Neale, et al. 2011; Tzeng, et al. 2009; Wu, et al. 2011].

Non-parametric methods, such as U-statistic-based methods, have shown great promise for high-dimensional data analysis, especially when the underlying phenotype distributions and modes of inheritance are unknown. Several U-statistic-based methods were recently adopted in genetic association studies to detect common variants underlying complex human diseases [Li 2012; Li, et al. 2011; Schaid, et al. 2005; Wei, et al. 2012; Wei, et al. 2013]. In this paper, we propose a non-parametric method, WU-SEQ, for testing the joint association of multiple SNVs with disease phenotypes. Built under the framework of the weighted U statistic, WU-SEQ is robust against different underlying distributions of phenotypes. As demonstrated by the simulation study, WU-SEQ had well-controlled type 1 errors and attained high power under binary, Gaussian, t-distributed and Cauchy phenotypes. In contrast, the performance of existing methods, such as SKAT, depends on satisfaction of the underlying assumptions. If the assumptions were violated (e.g., the distribution followed a heavy-tailed distribution such as Cauchy), SKAT had an inflated type 1 error, and had low power to detect an association. Although SKAT can handle classic types of phenotypes (e.g., the exponential family) by using appropriate link functions, WU-SEQ can be applied to a wider range of phenotypes without any distribution assumptions. In this paper, we used the normal quantile to build the test statistic. Alternatively, one can also use rank to construct the test statistic. When the sample size is sufficiently large, we expect that using quantile and using rank would have similar results. Nevertheless, for small sample sizes, using a normal quantile could be more powerful and less conservative than using rank. Yet, when the distribution is heavily tailed (e.g., Cauchy), using rank could attain a slight advantage than using quantile (e.g., having a well-controlled type I error) for small sample sizes (Table 6). As a similarity-based method, WU-SEQ is flexible, to accommodate various types of data. The genetic variants used to construct genetic similarity are not constrained to categorical data (e.g. SNV); they can also be count data (e.g., CNV) and continuous data (e.g., expression data). For high-dimensional sequencing data, the number of genetic variants evaluated in the association analyses can be large. We have showed in the simulation studies, when the number of variants was much larger than sample sizes, WU-SEQ could still control type I error and have good power performance (Table 3). Similar as existing methods (e.g., SKAT), WU-SEQ is gene/region based methods, and therefore can be directly applied to whole-exome sequencing data. Nevertheless, challenges remain when applying the method to whole-genome sequencing data because it requires a prior determination of a region or a functional unit (e.g., gene). In addition, potential issues, such as determining genome-wide significance, also need to be carefully considered for whole-genome sequencing data analysis. Although research has been initiated to address these issues[Xu, et al. 2014], further investigation is much needed on these topics.

In WU-SEQ, we first summarize the information from multiple markers into genetic similarities, and then evaluate the genetic similarities with the corresponding phenotypic similarities. If the genetic similarities are concordant with the trait similarities, we anticipate a large value of the test statistic, from which we could infer an association. In this paper, we use a weighted IBS to construct genetic similarity, which assigns more weights to rare variants. Other types of genetic similarity metrics can also be used, such as those consider interactions. Prior knowledge can also be incorporated, by assigning different weight for each variant according to their biological plausibility. To measure the phenotype similarity, we use the cross product kernel, $s_{i,i'} = h(q_i, q_{i'}) = q_i q_{i'}$, which lead to nice asymptotic properties of the test statistic. Nevertheless, we can use other types of kernel functions for the phenotype similarity. If the kernel function satisfies two regularity conditions, $E_F(h^2(Y_1, Y_2)) < \infty$ and $Var(E(h(Y_1, Y_2) | Y_2)) = 0$, we can calculate the asymptotic p-value by approximating the weighted U to a linear combination of chi-squared variables. Even if the two conditions are not satisfied, we can still use permutation to obtain the p-value of the association test. The choice of the genetic- or phenotype-similarity metrics depends on the underlying disease model, which could lead to different performance. For example, when the percentage of risk variants increases and/or when the effect sizes of risk variants follow one direction, the burden tests are expected to be more powerful than WU-SEQ. In this case, we can use genetic similarities accommodating the underlying disease models. One of the strategies is to first collapse all genetic variants by weighted sum and then calculate the genetic similarity.

Although joint association analyses greatly reduce the dimensionality for sequencing data, the computation could be intense if a permutation test is used. We derived the asymptotic distribution for WU-SEQ to facilitate the high-dimensional data analysis. When the sample size of a study is small, asymptotic properties may not hold and a permutation test can be used. We also use a projection method in WU-SEQ to take covariates into account. Through simulations and a real data analysis, we found the covariate-adjusted WU-SEQ was robust to different types of phenotype distributions. In addition, we found that WU-SEQ had almost the same power as SKAT when the phenotype followed Gaussian distribution and covariates were not considered. The reason is that SKAT uses a variance component score test, and WU-SEQ has a close connection with the variance component score test for the Gaussian-distributed phenotype. In fact, when covariates are not considered, SKAT can be viewed as a special case of WU-SEQ by using the cross product kernel without quantile transformation. When covariates are considered, especially when the phenotype does not follow the Gaussian distribution (i.e. the link function in SKAT is not an identity link), WU-SEQ could attain more computational efficiency than SKAT. In SKAT, one needs to first fit the null model using a generalized linear model, which involves iterative estimation. Furthermore, the calculation of a projection matrix in SKAT involves additional matrix multiplications (i.e., calculating $H = V - VX(X^T V X)^{-1} X^T V$ in SKAT vs. calculating $H = I - X(X^T X)^{-1} X^T$ in WU-SEQ, where $V$ is the covariance matrix estimated from the null model), and the calculation of the limiting distribution in SKAT involves an additional large matrix decomposition (i.e., calculating $H^{1/2} K H^{1/2}$, where $H^{1/2}$ need to be calculated from an eigen-decomposition of the $H$ matrix). The covariate-adjusted WU-SEQ does not require iterative estimation and needs less matrix multiplication and decomposition, which offers a greater computational advantage over SKAT.

In the analysis of the DHS study, WU-SEQ detected a strong association of *ANGPTL 4* with VLDL. By further studying the distribution of VLDL, we found the distribution was heavily skewed, which does not fit the normal assumption. The advantages of WU-SEQ are not limited to this specific case; the method could be useful for other cases (e.g., small sample sequencing studies and sequencing studies with extreme-phenotype design). A recent version of SKAT also considers the extreme phenotype by assuming a truncated Gaussian distribution[Barnett, et al. 2013]. While SKAT needs to make adjustments and certain assumptions for the truncated distribution of extreme phenotypes WU-SEQ, as a non-parametric method, can be directly applied to studies with extreme-value phenotypes.

**Appendices**

**Appendix A**

We first introduce a regularity condition on the weight function $k_{i,i'}$. Because $0 \leq w_{i,i'} \leq 1$ and $k_{i,i'} = w_{i,i'} - c$, we have $0 \leq |k_{i,i'}| \leq c'$, where $c'$ is a positive constant. Based on this, we have $\lim_{n \to \infty} \frac{1}{n(n-1)} \sum_{i \neq i'} k_{i,i'}^2 = C$, where $C$ is also a positive constant. Then, we can show,

$$E(\chi_{\Sigma,\infty}^2) = \sum_{l=1}^{\infty} \tilde{\lambda}_l E(\chi_{1,l}^2) = \sum_{l=1}^{\infty} \tilde{\lambda}_l = 0,$$

and,

$$\begin{aligned} Var(\chi_{\Sigma,\infty}^2) &= \sum_{l=1}^{\infty} \tilde{\lambda}_l^2 Var(\chi_{1,l}^2) \\ &= \lim_{n \to \infty} \frac{2}{(n-1)^2} \sum_{l=1}^{n} \lambda_l^2 \\ &= \lim_{n \to \infty} \frac{2}{(n-1)^2} trace(KK) \\ &= \lim_{n \to \infty} \frac{2}{(n-1)^2} \sum_{i \neq i'} k_{i,i'}^2 \\ &= 2C. \end{aligned}$$

Based on this result, we can conclude that $\chi_{\Sigma,\infty}^2 = \sum_{l=1}^{\infty} \tilde{\lambda}_l \chi_{1,l}^2$ has zero mean and finite variance. Additionally, the results implies that $\sqrt{n} WU_{seq}$ is degenerated, i.e., $\sqrt{n} WU_{seq} \xrightarrow{p} 0$.

**Acknowledgements**

We thank Xiaowei Zhan, Dajiang Liu, and Jonathan Cohen for helping us access the Dallas Heart Study dataset. This work was supported by the National Institute on Drug Abuse under Award Number K01DA033346 and by the National Institute of Dental & Craniofacial Research under Award Number R03DE022379.

**Tables:**

Table 1: Type I error and power comparisons of WU-SEQ and SKAT under different direction and magnitude of effects

| Effect* | Pct** | Binary | | Normal | | Student's t | | Cauchy | |
|---|---|---|---|---|---|---|---|---|---|
| | | SKAT | WU-SEQ | SKAT | WU-SEQ | SKAT | WU-SEQ | SKAT | WU-SEQ |
| **Type I Error** | | | | | | | | | |
| Null | 0 | 0.037 | 0.037 | 0.037 | 0.043 | 0.110 | 0.038 | 0.194 | 0.041 |
| **Power** | | | | | | | | | |
| A1 | 5 | 0.281 | 0.284 | 0.174 | 0.185 | 0.164 | 0.193 | 0.167 | 0.116 |
| | 10 | 0.417 | 0.423 | 0.315 | 0.329 | 0.212 | 0.331 | 0.160 | 0.201 |
| | 30 | 0.828 | 0.841 | 0.675 | 0.688 | 0.369 | 0.695 | 0.192 | 0.451 |
| | 50 | 0.935 | 0.944 | 0.856 | 0.869 | 0.525 | 0.880 | 0.177 | 0.635 |
| A2 | 5 | 0.065 | 0.073 | 0.085 | 0.093 | 0.140 | 0.134 | 0.190 | 0.095 |
| | 10 | 0.166 | 0.173 | 0.155 | 0.163 | 0.192 | 0.288 | 0.175 | 0.179 |
| | 30 | 0.539 | 0.547 | 0.524 | 0.527 | 0.471 | 0.784 | 0.199 | 0.578 |
| | 50 | 0.773 | 0.780 | 0.790 | 0.798 | 0.733 | 0.960 | 0.187 | 0.818 |

* "Null" corresponds to the null model with no functional variant; "A1" corresponds to the setting where half of the functional rare variants have deleterious effect and the other half of functional rare variants have protective effect; "A2" corresponds to the setting where all the functional rare variants have deleterious effects.
** Percentage of functional rare variants.

Table 2: Type I error and power comparisons of WU-SEQ and SKAT for different sample sizes

| Effect* | Sample size | Binary | | Normal | | Student's t | | Cauchy | |
|---|---|---|---|---|---|---|---|---|---|
| | | SKAT | WU-SEQ | SKAT | WU-SEQ | SKAT | WU-SEQ | SKAT | WU-SEQ |
| **Type I Error** | | | | | | | | | |
| Null | 50 | 0.001 | 0.005 | 0.027 | 0.047 | 0.117 | 0.044 | 0.118 | 0.041 |
| | 100 | 0.007 | 0.013 | 0.035 | 0.050 | 0.131 | 0.051 | 0.134 | 0.052 |
| | 200 | 0.022 | 0.029 | 0.029 | 0.039 | 0.121 | 0.043 | 0.165 | 0.056 |
| | 500 | 0.037 | 0.037 | 0.037 | 0.043 | 0.110 | 0.038 | 0.194 | 0.041 |
| **Power** | | | | | | | | | |
| A1 | 50 | 0.016 | 0.035 | 0.145 | 0.177 | 0.160 | 0.164 | 0.126 | 0.086 |
| | 100 | 0.119 | 0.181 | 0.279 | 0.318 | 0.220 | 0.296 | 0.160 | 0.164 |
| | 200 | 0.494 | 0.536 | 0.521 | 0.545 | 0.317 | 0.557 | 0.178 | 0.291 |
| | 500 | 0.935 | 0.944 | 0.856 | 0.869 | 0.525 | 0.880 | 0.177 | 0.635 |

* "Null" corresponds to the null model with no functional variant, "A1" corresponds to the setting where half of the functional rare variants have deleterious effect and the other half of functional rare variants have protective effect.

Table 3: Type I error and power comparisons of WU-SEQ and SKAT when number of variants is much larger than sample size*

| Effect | Method | Distribution | | | |
|---|---|---|---|---|---|
| | | Binary | Normal | Student's t | Cauchy |
| | | **Type I Error** | | | |
| Null | SKAT | 0.007 | 0.021 | 0.130 | 0.196 |
| | WU-SEQ | 0.012 | 0.059 | 0.042 | 0.059 |
| | | **Power** | | | |
| A1 | SKAT | 0.153 | 0.896 | 0.708 | 0.264 |
| | WU-SEQ | 0.304 | 0.933 | 0.864 | 0.558 |

*The average number of SNVs is 1873 in these settings, while the sample size is set as 100. The effect sizes is set as the same as in Table S2

Table 4: Type I error and power comparisons of WU-SEQ and SKAT for covariate adjustment

| Effect | Sample size | Binary | | Normal | | Student's t | | Cauchy | |
|---|---|---|---|---|---|---|---|---|---|
| | | SKAT | WU-SEQ | SKAT | WU-SEQ | SKAT | WU-SEQ | SKAT | WU-SEQ |
| | | **Type I Error** | | | | | | | |
| Null | 50 | 0.027 | 0.027 | 0.040 | 0.059 | 0.084 | 0.069 | 0.128 | 0.063 |
| | 100 | 0.030 | 0.034 | 0.042 | 0.054 | 0.106 | 0.064 | 0.131 | 0.065 |
| | 200 | 0.028 | 0.028 | 0.036 | 0.050 | 0.110 | 0.060 | 0.173 | 0.061 |
| | 500 | 0.036 | 0.036 | 0.043 | 0.048 | 0.115 | 0.065 | 0.194 | 0.047 |
| | | **Power** | | | | | | | |
| A1 | 50 | 0.060 | 0.071 | 0.128 | 0.165 | 0.297 | 0.400 | 0.140 | 0.112 |
| | 100 | 0.156 | 0.210 | 0.306 | 0.342 | 0.483 | 0.664 | 0.164 | 0.159 |
| | 200 | 0.522 | 0.545 | 0.500 | 0.527 | 0.682 | 0.882 | 0.163 | 0.262 |
| | 500 | 0.929 | 0.934 | 0.850 | 0.855 | 0.884 | 0.997 | 0.180 | 0.554 |

* "Null" corresponds to the null model with no functional variant; "A1" corresponds to the setting where half of the functional rare variants have deleterious effect and the other half of functional rare variants have protective effect.

Table 5: Type I error and power comparisons of WU-SEQ by using different $c$*

| Effect | Method | Distribution | | | |
|---|---|---|---|---|---|
| | | Binary | Normal | Student's t | Cauchy |
| | | **Type I Error** | | | |
| Null | WU-SEQ$_{L1}$** | 0.013 | 0.048 | 0.056 | 0.041 |
| | WU-SEQ$_{L2}$*** | 0.013 | 0.049 | 0.058 | 0.043 |
| | | **Power** | | | |
| A1 | WU-SEQ$_{L1}$ | 0.164 | 0.313 | 0.300 | 0.150 |
| | WU-SEQ$_{L2}$ | 0.170 | 0.322 | 0.306 | 0.159 |

*The sample size for this simulation is 100 and the effect sizes is set as the same as in Table S2.
**In WU-SEQ$_{L1}$, we choose c by using L1 norm.
***In WU-SEQ$_{L2}$, we choose c by using L2 norm.

Table 6: Type I error and power comparison of WU-SEQ by using normal quantile or rank*

| Effect | Method | Distribution | | | |
|---|---|---|---|---|---|
| | | Binary | Normal | Student's t | Cauchy |
| **Type I Error** | | | | | |
| Null | WU-SEQ$_{RK}$** | 0.017 | 0.02 | 0.031 | 0.032 |
| | WU-SEQ$_{QT}$*** | 0.017 | 0.047 | 0.057 | 0.050 |
| **Power** | | | | | |
| A1 | WU-SEQ$_{RK}$ | 0.161 | 0.243 | 0.304 | 0.181 |
| | WU-SEQ$_{QT}$ | 0.161 | 0.314 | 0.322 | 0.169 |

*The sample size for this simulation is 100 and the effect sizes is set as the same as in Table S2.
**In WU-SEQ$_{RK}$, we use cross product kernel based on rank of the phenotype value without quantile transformation.
***In WU-SEQ$_{QT}$, we use cross product kernel with quantile transformation.

Table 7: The association of 4 candidate genes with 3 continuous phenotypes (i.e., BMI, Cholesterol, and VLDL) in the Dallas Heart Study

| Gene | BMI | | Cholesterol | | VLDL | |
|---|---|---|---|---|---|---|
| | SKAT | WU-SEQ | SKAT | WU-SEQ | SKAT | WU-SEQ |
| *ANGPTL3* | 0.633 | 0.752 | 0.559 | 0.662 | 0.471 | 0.198 |
| *ANGPTL 4* | 0.121 | 0.255 | 0.16 | 0.316 | 0.105 | **0.007** |
| *ANGPTL 5* | 0.633 | 0.607 | 0.95 | 0.926 | 0.683 | 0.664 |
| *ANGPTL 6* | 0.874 | 0.773 | **0.025** | 0.059 | 0.433 | 0.453 |
| *All 4 genes* | 0.503 | 0.641 | 0.117 | 0.373 | 0.316 | **0.032** |

* P-value from the association analysis, adjusting for age, gender, and race.

Figure 1: Distribution of SNVs with MAF<0.03 for the 4 genes in the Dallas Heart Study

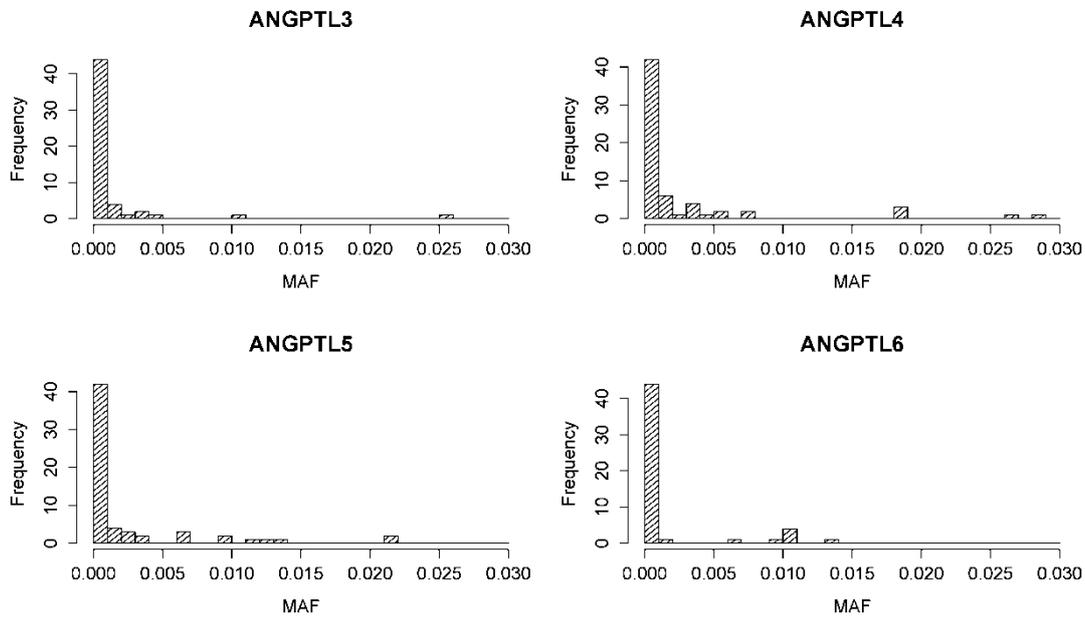

Figure 2: Distributions of the 3 phenotypes in the Dallas Heart Study

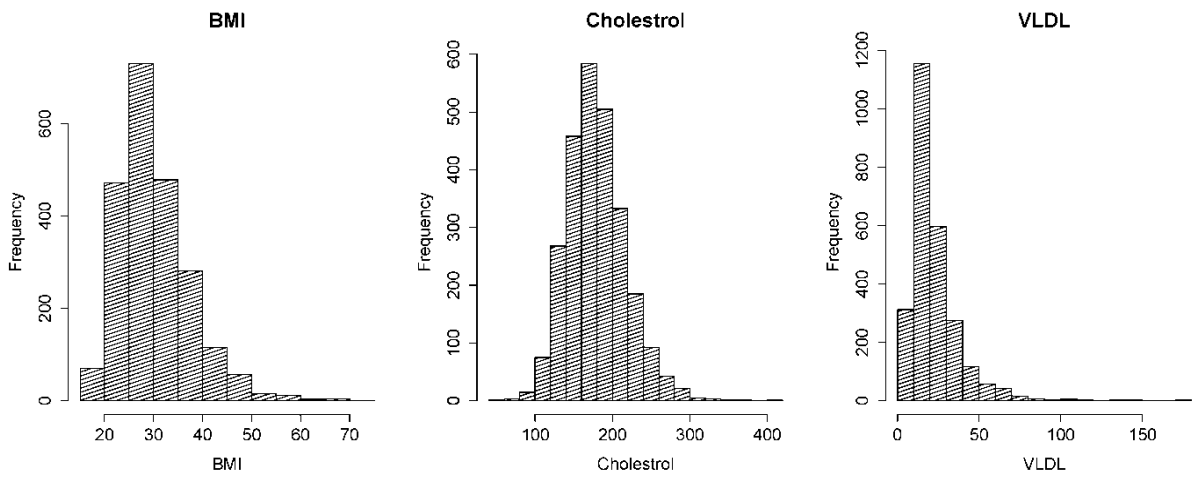